\documentclass[aps,prl,twocolumn,superscriptaddress,showpacs]{revtex4-1}

\usepackage{hyperref}
\usepackage{graphicx}
\usepackage{epsfig}
\usepackage{subfigure}
\usepackage{setspace}
\usepackage{amsmath}

\begin{document}

\title{Monitoring the formation of oxide apertures in micropillar cavities}
\author{Morten P. Bakker}
\affiliation{Huygens Laboratory, Leiden University, P.O. Box 9504, 2300 RA Leiden, The Netherlands}
\author{Henk Snijders}
\affiliation{Huygens Laboratory, Leiden University, P.O. Box 9504, 2300 RA Leiden, The Netherlands}
\author{Donald J. Suntrup III}
\affiliation{University of California Santa Barbara, Santa Barbara, California 93106, USA}
\author{Tuan-Ahn Truong}
\affiliation{University of California Santa Barbara, Santa Barbara, California 93106, USA}
\author{Pierre M. Petroff}
\affiliation{University of California Santa Barbara, Santa Barbara, California 93106, USA}
\author{Martin P. van Exter}
\affiliation{Huygens Laboratory, Leiden University, P.O. Box 9504, 2300 RA Leiden, The Netherlands}
\author{Dirk Bouwmeester}
\affiliation{Huygens Laboratory, Leiden University, P.O. Box 9504, 2300 RA Leiden, The Netherlands}
\affiliation{University of California Santa Barbara, Santa Barbara, California 93106, USA}

\date{\today}
\begin{abstract}

An imaging technique is presented that enables monitoring of the wet thermal oxidation of a thin AlAs layer embedded between two distributed Bragg reflector (DBR) mirrors in a micropillar.
After oxidation we confirm by white light reflection spectroscopy that high quality optical modes confined to a small volume have been formed.
The combination of these two optical techniques provides a reliable and efficient way of producing oxide apertured micropillar cavities for which the wet thermal oxidation is a critical fabrication step.

\end{abstract}

\maketitle

Wet thermal oxidation is a widely used technique that allows for optical (electrical) confinement by making use of the large difference in effective refractive index (resistivity) between oxidized and unoxidized regions\cite{Baca2005}.
It is an important fabrication step in the production of vertical cavity surface emitting lasers (VCSELs) \cite{Wiedenmann2006} and of certain optical microcavities \cite{Stoltz2005}. 
Embedding self-assembled quantum dots (QDs) in such microcavities enabled the study of cavity quantum electrodynamics in solid state systems \cite{Rakher2009}.
Such devices are of interest for single-photon sources \cite{Strauf2007} and for hybrid quantum-information devices \cite{Bonato2010}.

A real challenge, however, is to obtain a high fabrication reliability in the wet oxidation process, since the oxidation rate depends strongly on parameters which are difficult to reproduce, such as the roughness of the etched sidewalls or the Al composition to within an $\sim1\%$ accuracy.
As repetitive calibrations can be costly, techniques to monitor the oxide formation in situ are important. All three techniques reported so far make use of the change in the reflectivity spectrum as one or more Al containing layers are being oxidized.
The average change in reflectivity of a large patterned area  can be monitored \cite{Sakamoto2002}. The reflectivity contrast between a fully oxidized DBR mirror and unoxidized areas can be directly viewed through a charge coupled device (CCD) camera using broadband light \cite{feld1998}. Finally a buried, thin oxide aperture layer can be imaged \cite{Almuneau2008}, but this technique is delicate as the applied wavelength interval has to be tuned carefully such that the reflectivity contrast allows for sufficient spatial resolution.

\begin{figure}[b]
\centering
    \centerline{\includegraphics[angle=0]{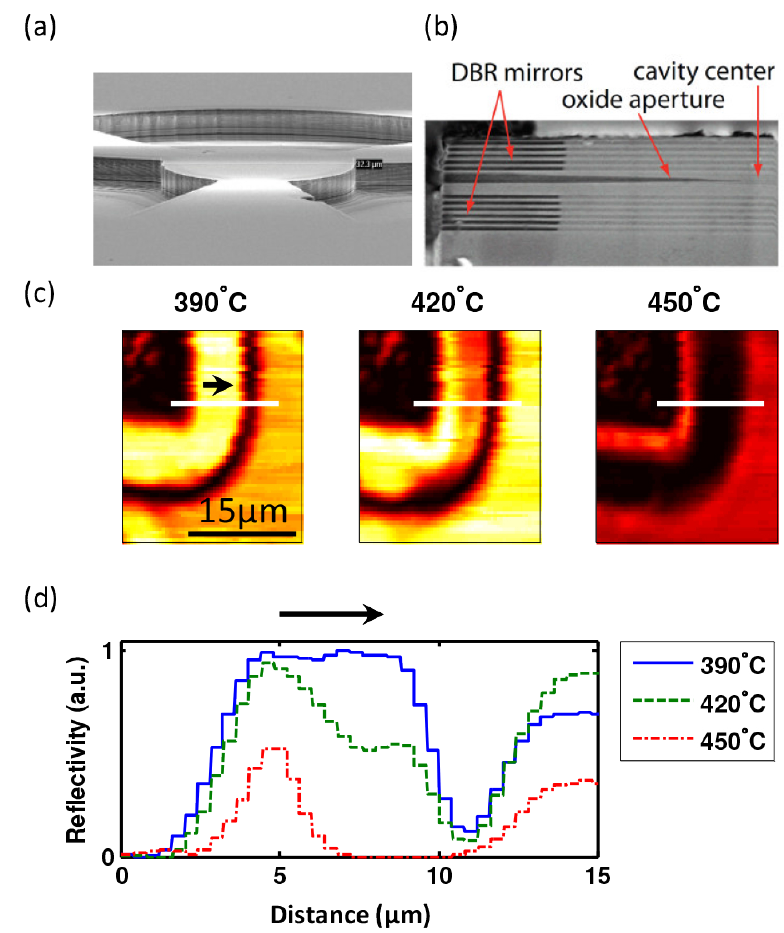}}
    \caption{(a) SEM image of the three etched trenches that form a micropillar mesa connected to the bulk material. (b) SEM cross-sectional image of the center of a micropillar. Al$_{x}$O$_{y}$ is darker than the Al$_x$Ga$_{1-x}$As layers. (c) Reflection scans (light: high  reflectivity) taken with a 1064 nm laser for different temperatures at the corner of an etched trench (white box in Fig. \ref{LgAreaScan} shows  this corner for a different micropillar). (d) Reflectivity as a function of the distance along the white lines in (c).}
    \label{Tempscan}
\end{figure}

In this paper, we present a new imaging technique based on measuring the reflectivity, while scanning with a monochromatic beam over an etched micropillar structure. We use a wavelength on the red side of the photonic stopband, such that absorption in the semiconductor material is low for temperatures up to 450$^{\circ}$C. For an oxidized sample we are able to clearly identify three regions: unoxidized regions, regions where only a thin aperture layer has been oxidized and regions where both the aperture region and the DBR mirror are oxidized.
We then monitor a sample while it is being oxidized and stop the oxidation when only a small unoxidized area is left at the center of a micropillar structure. 
After oxidation, we perform an in situ characterization at room temperature of the optical modes defined by the optically confining oxide aperture.

\begin{figure*}[t!]
\centering
    \centerline{\includegraphics[ angle=0]{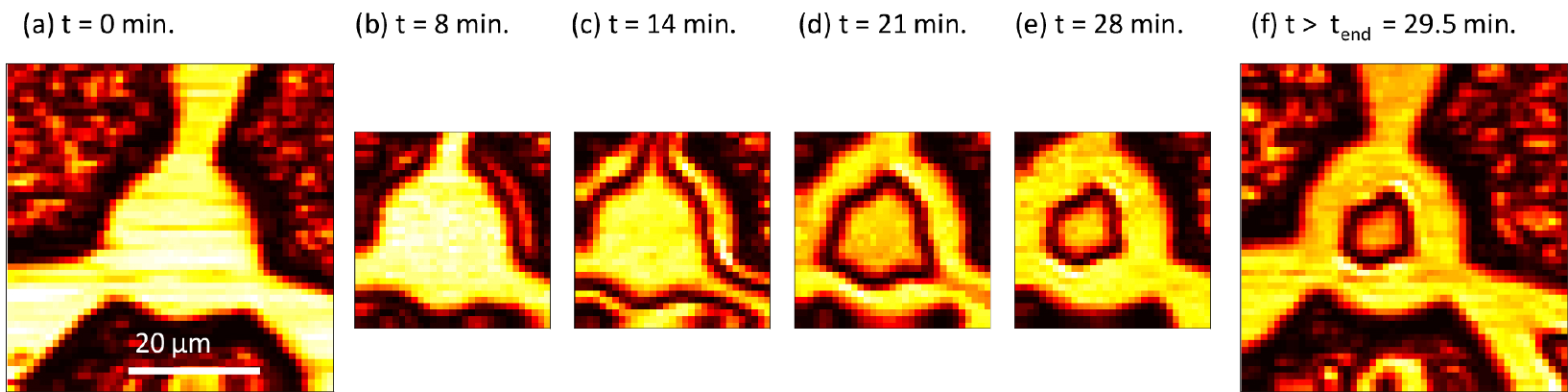}}
    \caption{(a) Reflectivity scan (light: high reflectivity) of a 1064 nm laser on an etched micropillar before the oxidation was started. (b-e) Reflectivity scans for different times after the oxidation was started. Every scan takes about 1.5 minutes. (f) Reflectivity scan taken after the oxidation was stopped after 29.5 minutes.}
    \label{OxMov}
\end{figure*}

The sample is grown by molecular beam epitaxy on a GaAs [100] substrate and is constructed as follows: two DBR mirrors, comprising alternating $\lambda/4$-thick layers of GaAs and Al$_{0.9}$Ga$_{0.1}$As, embed an aperture layer, consisting of a thin 10 nm AlAs layer in between Al$_{0.75}$Ga$_{0.25}$As and Al$_{0.83}$Ga$_{0.17}$As, and an active layer containing InGaAs/GaAs self-assembled quantum dots inside a PIN diode structure.
Trenches are etched down to the bottom DBR using reactive ion etching, leaving a 30 $\mu$m circular micropillar connected by three narrow unetched channels to the bulk material (see Fig. \ref{Tempscan}a). Global electrical contacts to the doped layers allow for control of the electrical field inside every micropillar in an array of 42 micropillars.

The conversion of Al$_{x}$Ga$_{1-x}$As into Al$_{x}$O$_{y}$ takes place by applying H$_2$O vapor to the sample at a typical process temperature of 400-450$^{\circ}$C.
Special care is taken to prevent dry oxidation with oxygen, as it has been reported that this can prevent further wet oxidation \cite{Baca2005}.
First native oxide is removed from the Al$_{x}$Ga$_{1-x}$As by dipping the sample for 20s in NH$_4$OH, then it is rinsed with demi-water and placed in isopropanol.
The sample is finally clamped to a holder, equipped with a heater and a thermocouple, inside an oxidation chamber fitted with a viewport. Isopropanol is maintained on the sample surface until the chamber has been filled with nitrogen gas, in order to prevent exposure to oxygen.
Steam is applied by flowing nitrogen carrier gas through a water-filled bubbler heated to 95$^{\circ}$C.

In Fig \ref{Tempscan}(b) a SEM cross sectional image shows that the oxide penetration depth varies for different layers. Our goal is to view and monitor the growth of the oxidized layers in situ with a non-invasive optical technique.


\begin{figure}
\centering
    \centerline{\includegraphics[angle=0]{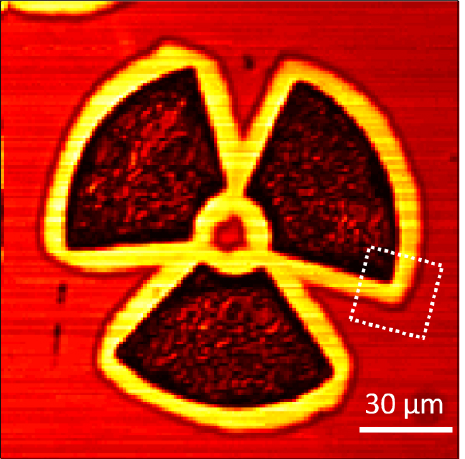}}
    \caption{Reflectivity scan (light: high reflectivity) of the same micropillar as in Fig. \ref{OxMov}, taken over a larger region at room temperature.}
    \label{LgAreaScan}
\end{figure}

A NA=0.40 and 20 mm working distance objective is used to image the sample surface on a CCD camera and to focus incident light to a diffraction limited spot size.
In order to differentiate between regions with a different extent of oxidation, we monitor the reflectivity of a focused Nd:YAG 1064 nm laser beam that we scan across the sample.
In order to find the operating temperature that yields the best reflectivity contrast, we take spatial reflectivity scans of an already oxidized sample at the edge of an etched trench (see Fig \ref{Tempscan}(c)). The reflectivity change with refractive index is temperature dependent.
In Fig. \ref{Tempscan}(d) the reflection intensities along the white lines in Fig \ref{Tempscan}(c) are presented.
Starting from the left side of the graph where the etched trench is, the reflectivity rapidly changes from low to high when the oxidized DBR is reached.
In the middle a constant reflectivity is visible, corresponding to an oxide layer with a constant thickness. On the right side the gradient in the oxide layer, at the end of the oxide aperture, causes a varying reflectivity after which the reflectivity of the unoxidized region is more or less constant.

When performing a monitored oxidation with this specific wafer material, a temperature of 420$^{\circ}$C is chosen. At this temperature the contrast between the oxide aperture and the unoxidized regions is larger than at higher temperature, and the oxidation rate is fast enough to complete the oxidation in about 30 minutes.
In order to monitor the oxidation at a different temperature with similar contrast, a light source with a different wavelength could be used.


First we take an unoxidized sample and perform a spatial reflectivity scan of the micropillar center. Fig. \ref{OxMov} (a) shows a homogeneous reflectivity for the unetched regions and a lower reflectivity for the etched parts.
Once the oxidation process is started, continuously scans are recorded, which take about 1.5 minutes each. Several of these scans are shown in Fig \ref{OxMov} (b-e).

The front of the oxide aperture, in this case characterized by a lower reflectance, starts to penetrate towards the micropillar center. After 29.5 minutes the oxide taper has advanced to within 3 $\mu$m of the pillar center and we stop the oxidation by shutting off the water vapor flow and purging the oxidation chamber with nitrogen.
Several minutes after the oxidation was stopped a final scan is taken (Fig. \ref{OxMov} (f)).

\begin{figure}[t]
\centering
    \centerline{\includegraphics[angle=0]{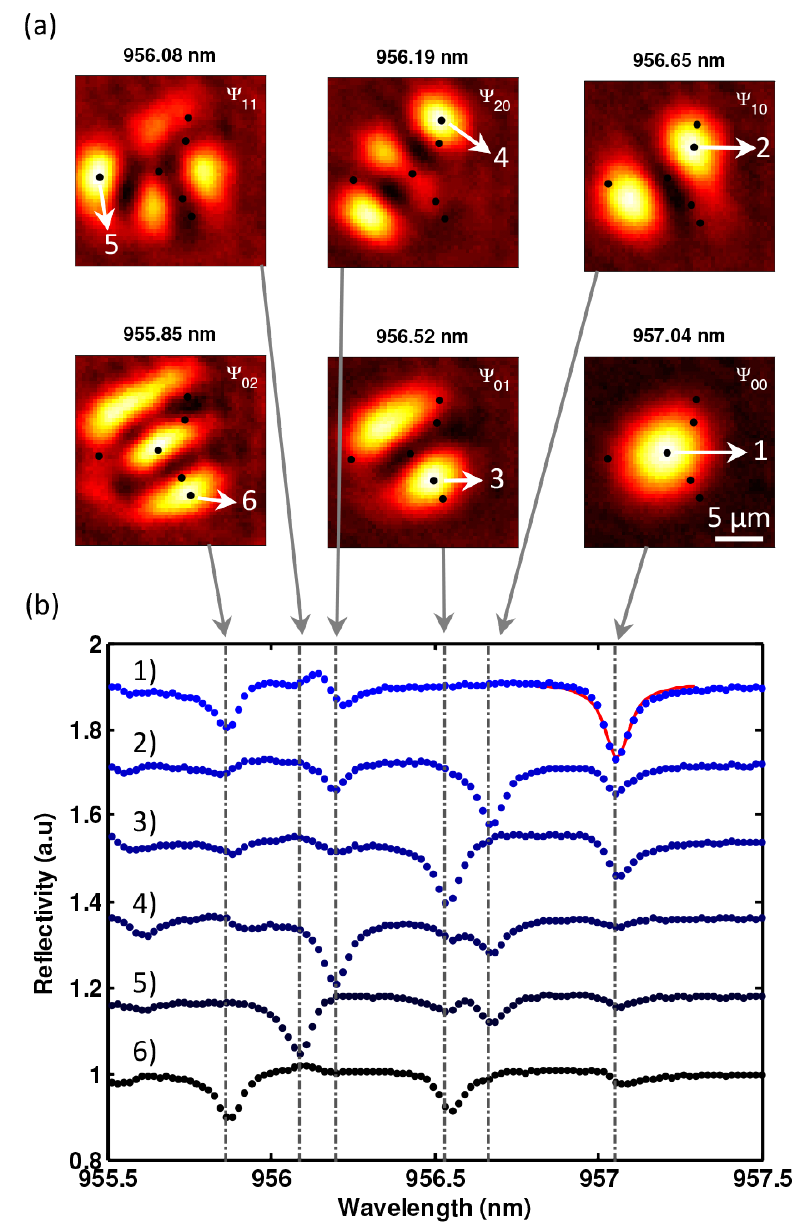}}
    \caption{(a) Reflectivity scans (light: lower reflectivity) for the indicated wavelengths, between 955.5-957.5 nm, as a function of position in the center of a micropillar. (b) Reflection spectra at the six positions that are marked with a black dot and labelled with numbers 1-6. An offset is added between each trace. The red line in curve 1 is a fit to determine the $Q$-factor of the $\Psi_{00}$ mode.}
    \label{ModeChar}
\end{figure}

Once the sample has cooled down to room temperature, a large spatial reflectivity scan is taken, see Fig. \ref{LgAreaScan}. The unoxidized center, having the same reflectivity as regions far away from the etched trenches, is seen to have the same size and shape as the center of the scan in Fig. \ref{OxMov}(f), which was recorded at 420$^{\circ}$C.

In order to characterize the confined optical modes of the device, the center of the micropillar is illuminated using a superluminescence diode (Superlum, 920-980nm), with a wavelength range covering the optical modes, and the reflection spectra are recorded with a spectrometer as a function of position \cite{Ctistis2010}.
Fig. \ref{ModeChar}(a) shows spatial scans of the reflected intensity at the resonance wavelengths of the first six Hermite-Gaussian modes.
Fig. \ref{ModeChar}(b) shows the reflection spectra at six positions, marked in Fig. \ref{ModeChar}(a) with black dots and numbers. These positions correspond to the minima in the reflection spectra and demonstrate a clear spatial separation of the different modes.
In order to determine the $Q$-factor, we fit a Lorentzian reflection dip to the $\Psi_{00}$ mode and find a $Q$-factor 10,000-12,000.
For the average wavelength splitting $\Delta\lambda$, between the fundamental $\Psi_{00}$ and first order $\Psi_{10}$ and $\Psi_{01}$ modes, we find $\Delta\lambda \approx 0.62$ nm.

An important figure of merit of microcavities is the Purcell factor $P = \frac{3}{4\pi^2}(\frac{\lambda_0}{n})^3\frac{Q}{V}$, where $\lambda_0$ is the cavity wavelength, $n$ is the refractive index, $Q$ is the quality factor and $V$ is the mode volume.
Assuming there is a quadratic confining potential present, giving rise to Hermite-Gaussian optical modes, this can be rewritten as: $P = \frac{12}{\pi}F(\frac{\Delta\lambda}{\lambda_0})$ \cite{Bonato2012}.
Here $F = \frac{\lambda_0 Q}{2nL_{eff}}$ is the cavity finesse and $L_{eff}$ is the effective cavity length.
By combining the measured $Q$-factor 10,000-12,000, $\Delta\lambda \approx 0.62$ nm with $n\approx3$ and $L_{eff} \approx 1.4$ $\mu$m estimated from the device structure, we find an estimated Purcell factor $P = 2.8-3.4$.

Higher Purcell factors can be obtained by performing the oxidation longer and thereby reducing the cavity volume. Also the $Q-$factor, limited by absorption in the doped layers, can be further increased by cooling the device to a temperature of 4K, where such micropillars containing quantum dots are typically operated.
The difference in the wavelength splitting, between the fundamental $\Psi_{00}$ and either one of the first order $\Psi_{10}$ and $\Psi_{01}$ modes, is a result of elipticity arising from preferential oxide formation rates and can be compensated for by etching elliptical micropillars.
Finally we would like to note that the fundamental cavity mode in micropillars is typically polarization non-degenerate due to birefringence, although we were limited by the resolution of the spectrometer of 0.019 nm and could not characterize this. Permanent tuning to compensate for this birefringence has been demonstrated \cite{Bonato2009} and could be used as a final step in the fabrication process.


In conclusion, we have developed a procedure that allows for the fabrication and characterization with high accuracy of high quality micropillar cavities that, operating at cryogenic temperatures, are of interest for quantum information applications.

This work was supported by NSF under Grant No. 0901886, the Marie-Curie Award No. EXT-CT-2006-042580 and FOM-NWO Grant No. 08QIP6-2. We thank Cristian Bonato for useful discussions.


\bibliography{mortenpbakker}
\end{document}